\documentclass[%
aps,
jmp,%
amsmath,amssymb,
preprint,%
]{revtex4-2}

\usepackage{graphicx}
\usepackage{dcolumn}
\usepackage{bm}

\usepackage{color}
 \usepackage{multirow}
 \usepackage{rotating}
 \usepackage{amsmath}

\input{epsf} 

\begin{document}
\title{Simple but not Simpler: A Surface-Sliding Method for Finding the Minimum Distance between Two Ellipsoids}
\author{Dariush Amirkhani and Junfeng Zhang\footnote{
Corresponding author: Dr. Junfeng Zhang, School of Engineering and Computer Science, 
Laurentian University, 935 Ramsey Lake Road, Sudbury, ON P3E 2C6, Canada. 
Tel: 1-705-675-1151 ext. 2248; Email: jzhang@laurentian.ca.}
}
\affiliation{
School of Engineering and Computer Science, Laurentian University, 935 Ramsey Lake Road, Sudbury, Ontario, P3E 2C6, Canada 
}

\date{\today}

\begin{abstract}
		We propose a novel iterative process to 
		establish the minimum separation between two ellipsoids. 
		The method maintains one point on each surface and updates their locations in the $\theta-\phi$
		parametric space.
		The  {\it tension} along the connecting segment between the two surface points serves as 
		the guidance for the sliding direction and the distance between them thus decreases gradually.
		The minimum distance is established when the connecting segment becomes perpendicular to the ellipsoid surfaces, 
		whereas the net effect of the segment tension disappears and the surface points are not moving anymore.
		Demonstration examples are carefully designed, 
		and excellent numerical performance is observed, including accuracy, consistency, stability and robustness.
		Furthermore, compared to other existing techniques,
		this surface-sliding approach has several attractive features,
		such as the clear geometric representation, the concise formulation, 
		the simple algorithm and the potential to be extended to other situations straightforwardly.  
		This method is expected to be useful for future  studies in  computer graphics, engineering design, material modeling, and scientific simulations. 
		
~~
\end{abstract}

\maketitle
\newpage
\section{Introduction}
\label{sect:intro}

Finding the minimum distance between two ellipsoids in three-dimensional (3D) space is frequently encountered
in various engineering and research computations, such as collision and contact detection in robotics,  particle-resolved simulations of suspension flows, material modeling and computer graphics \cite{girault2022,Porous_model, Mayssaa_2023, compisite_2025,Porous_modeling_2022, Banna_2025,computer_graphics}.
In such applications, one needs not only the scalar separation distance but also the associated closest points and contact direction for collision forces or near-field lubrication interactions \cite{girault2022, Vincent_2021_Fluids,Mayssaa_2022, YZS_2025}.

A variety of algorithms have been proposed for this geometric query, 
and these methods have been recently reviewed by  Girault et al. \cite{girault2022} and Banna \cite{Banna_2023_phd}.
A widely used geometric approach is the moving-ball algorithm, 
which iteratively constructs interior tangent spheres and updates surface points as the intersection points of the line segment connecting the sphere centers with ellipsoid surfaces \cite{linhan2001}. 
The appeal lies in its conceptual simplicity, however, slow convergence has been noticed for small, thin ellipsoids  \cite{linhan2001,girault2022}.
On the other hand, 
convex-geometry approaches, instead, reduce the distance computation to a distance-to-origin problem for the Minkowski difference of two convex sets, leading to procedures such as the Gilbert-Johnson-Keerthi  algorithm \cite{gilbert1988,vandenbergen1999,girault2022}. 
These methods require support mappings and a simplex-update sub-algorithm, 
and recovering the closest points on the original ellipsoids involves an additional reconstruction step \cite{girault2022}.
Other formulations include the physics-inspired dynamics such as the Newton--Coulomb (charged-body) method \cite{abbasov2015} and the optimization-based methods such as the exact exterior penalty formulations \cite{tamasyan2014}. 
Comparative tests indicate that these methods can introduce nontrivial parameter calibration, overlap-prechecks, or substantial computational overhead, and they may be less attractive for large numbers of repeated distance queries in simulations \cite{girault2022}.
Furthermore, 
algebraic approaches based on Lagrange multipliers and elimination theory 
reduce the constrained minimum-distance problem to solving high-degree  polynomial systems \cite{uteshev2008}. 
While mathematically rigorous, comparative discussions emphasize that such implementations can be laborious and challenging to deploy robustly in practice \cite{girault2022,tamasyan2014}, and solving high-degree polynomial root problems can be numerically delicate in floating-point arithmetic \cite{higham2002}.
More recently, a machine learning approach has also been developed for estimating the distance between ellipsoids \cite{Banna_2023_phd}.

Regardless of the mathematical formulations and computational routes, the minimum-distance configuration satisfies a simple geometric characterization: 
when a pair of closest points exists, 
the segment joining the closest points aligns with the outward normals at both surface points \cite{linhan2001,girault2022}. 
Inspired by the moving-ball method \cite{linhan2001},
in this work, 
we propose an alternative to find the minimum distance and the corresponding locations on two ellipsoids
by allowing the two points, one on each ellipsoid, 
to slide smoothly on the surfaces under the direction of the connecting segment.
Compared to other existing methods in the literature,
no high-degree polynomial solver is needed, and 
the segment--surface intersection calculations in the moving-ball method are avoided.
The algorithm is simple,  straightforward and easy to implement, 
and explicit information is  readily  available during the search iteration process without reconstruction.
Numerical tests demonstrate the calculation is efficient, accurate and stable.
These features are valuable in situations, such as particle-resolved simulations,
where frequent and extensive calculations are needed for ellipsoid clusters.

The rest of this paper is organized as follows.
In Sect. \ref{sect:theory}, we first present the relevant mathematical formulations,
and then describe the surface-sliding iteration step in details.
Several numerical examples are provided in Sect. \ref{sect:demos}
to demonstrate the algorithm performance in accuracy, robustness and spatial isotropy.
Finally, summary and concluding remarks are presented in Sect. \ref{sect:summary}.

\begin{figure}[thb]
	\centering
	\includegraphics[width=0.5\textwidth, clip]{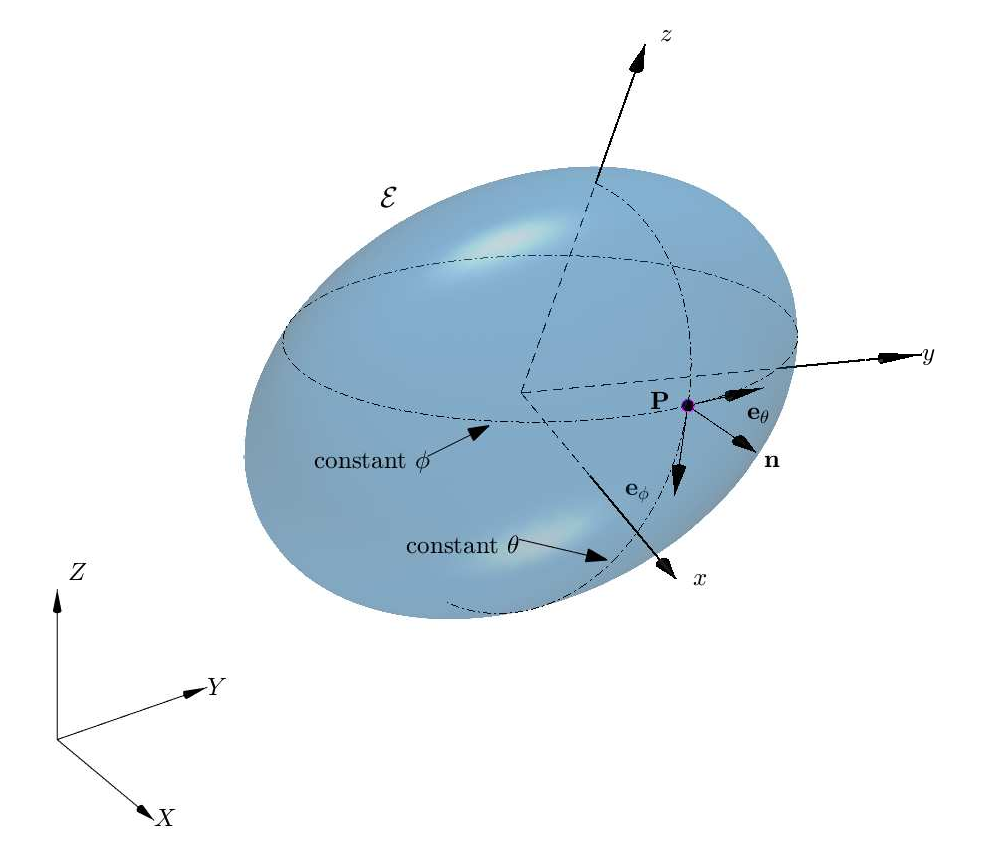}
	\caption{A graphic representation for the notations adopted in this article for an ellipsoid.}
	\label{fig:coordinates}
\end{figure}

\section{Method Description}
\label{sect:theory}

\subsection{Coordinate Systems and Ellipsoid Representation}
We first consider one ellipsoid $\mathcal{E}$ in the global coordinate ${\bf X}=[X, Y, Z]^T$ 
in Fig. \ref{fig:coordinates}.
A local coordinate system ${\bf x}=[x,y,z]^T$ is also established at the ellipsoid center ${\bf X}_0$ and along its axes.
The semi-axis lengths are denoted by $a$, $b$ and $c$, respectively, along the local coordinate directions 
$x$, $y$ and $z$. 
The transformation  between the global and local coordinates for a position 
can be expressed by
\begin{equation}
	{\bf X}={\bf R} {\bf x} +{\bf X}_0 ~~~,
	\label{eq:X_x}
\end{equation}
where ${\bf R}$ is the rotation matrix given as 
\begin{equation}
		\mathbf{R}=    
	\begin{bmatrix}
		1 & 0 &0\\
		0 & \cos\alpha &-\sin\alpha\\
		0 & \sin\alpha & \cos\alpha
	\end{bmatrix}
		\begin{bmatrix}
		\cos\beta &0&\sin\beta\\
		0&1&0\\
		-\sin\beta&0&\cos\beta
	\end{bmatrix}
	\begin{bmatrix}
		\cos\gamma &-\sin\gamma&0\\
		\sin\gamma&\cos\gamma&0\\
		0&0&1
	\end{bmatrix}~~~.
	\label{eq:RzRyRx_main}
\end{equation}
Here $\alpha$, $\beta$ and $\gamma$ are the Euler angles,
which describe the relative orientation between the global and local coordinate directions \cite{Euler_angle}.

The local coordinates of a point ${\bf P}$ on the ellipsoid surface $\mathcal{E}$ can be conveniently expressed 
using the parametric representation 
\begin{equation}
	{\bf x}=\left[ 
		a\sin\phi \cos\theta, ~
		b \sin\phi \sin \theta, ~
		c \cos \phi \right]^T ~~~,
	\label{eq:local_x}
\end{equation}
where $\phi \in [0, \pi]$ is the polar angle measured from the $z$ axis, and 
the parameter $\theta$ varies between 0 and $2 \pi$.
Therefore, any location on the surface $\mathcal{E}$ can be uniquely defined by a pair of these two parameters ($\theta, \phi$). 
Furthermore, normal (${\bf n}$) and tangent (${\bf e}_\theta$ and ${\bf e}_\phi$) directions at point 
${\bf P}$ can also be calculated from them as
 \begin{equation}
 	{\bf n}=\frac{\bf N}{|{\bf N}|}~,~~ 
 	{\bf N}= \left[ bc \sin \phi \cos \theta,~ 
 		ac \sin \phi \sin \theta,~
 		ab \cos \phi \right]^T~~~;
 	\label{eq:n}
 \end{equation}
 \begin{equation}
{\bf e}_\theta=\frac{{\bf r}_\theta} {|{\bf r}_\theta|}~,~~ 
{\bf r}_\theta=\left[ 
	-a\sin \phi \sin \theta,~ 
	b \sin \phi \cos \theta,~
	0 \right]^T~~~;
\label{eq:e_theta}
 \end{equation}
 \begin{equation}
 	{\bf e}_\phi=\frac{{\bf r}_\phi} {|{\bf r}_\phi|}~,~~ 
 	{\bf r}_\phi=\left[
 			a\cos \phi \cos \theta,~
 		b \cos \phi \sin \theta,~
 		-c\sin \phi \right]^T~~~.
 	\label{eq:e_phi}
 	 \end{equation}
All these unit vectors  are given here in the local coordinates.
The conversion of a vector ${\bf v}$ in the local coordinates to its counterpart ${\bf V}$ in the global coordinates can be easily done via the rotation matrix ${\bf R}$ as ${\bf V}={\bf R} {\bf v}$.
The tangential vector ${\bf e}_\theta$ indicates the direction along which 
parameter $\theta$ increases with angle $\phi$ unchanged;
and similarly ${\bf e}_\phi$ points to the direction of increasing $\phi$ at constant $\theta$.

\begin{figure}[thb]
	\centering
	\includegraphics[width=0.5\textwidth, clip]{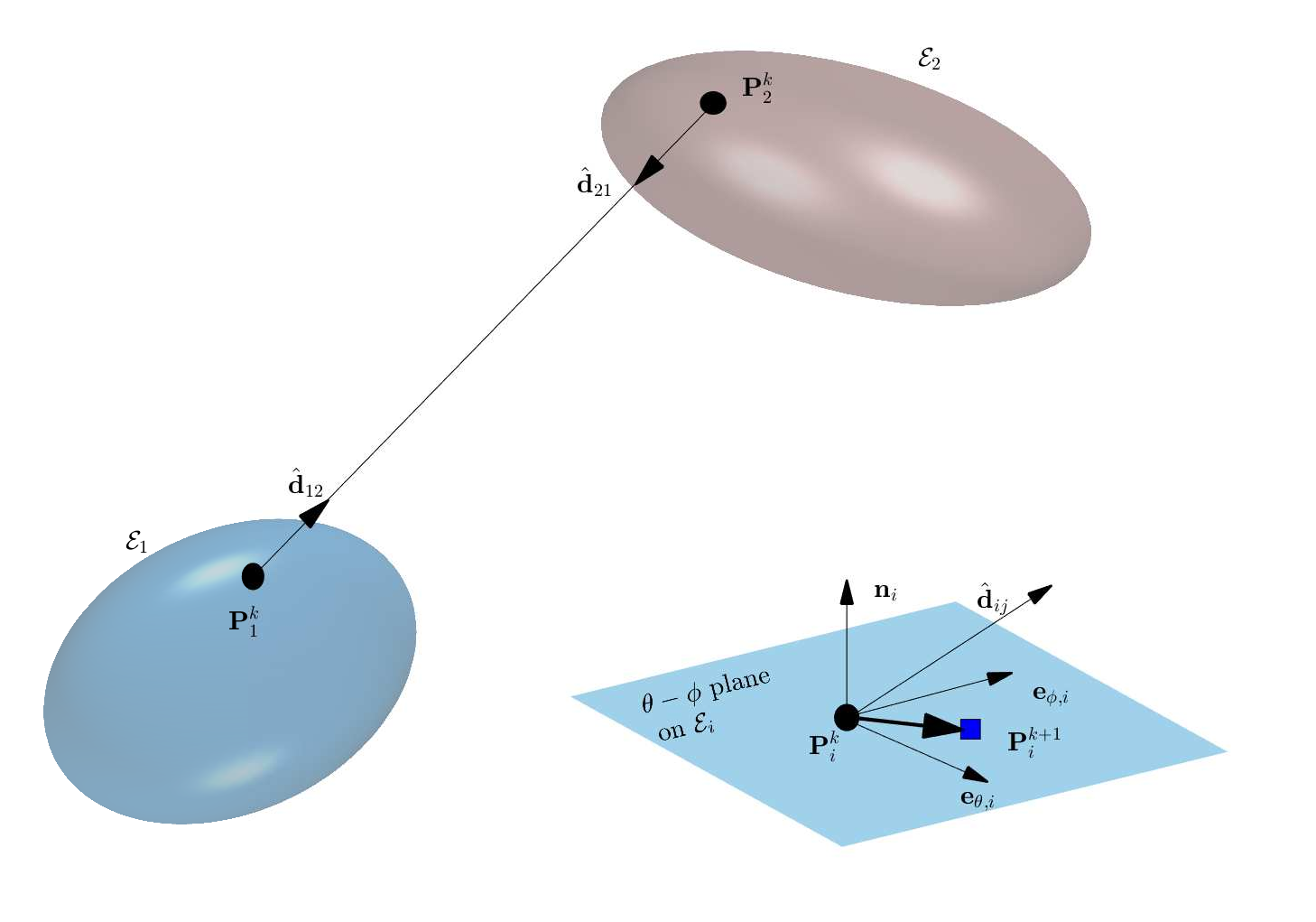}
	\caption{A graphic representation of the search process for the minimum distance between two ellipsoids, with key vectors used in the calculation illustrated.
		}
	\label{fig:two_particles}
\end{figure}

 \subsection{Update Process of Surface Point Locations}
 Now we consider the minimum distance problem for the two ellipsoids $\mathcal{E}_1$ and $\mathcal{E}_2$ shown in Fig. \ref{fig:two_particles}.
 The surface-sliding method involves an iterative process, 
 during which the two surface points, ${\bf P}_1$ on $\mathcal{E}_1$ and  ${\bf P}_2$ on $\mathcal{E}_2$,
 update their positions simultaneously to reduce the segment distance between them.
 For convenience, hereafter, we use the subscript  to indicate the ellipsoid and the superscript for the iteration step.  
For example,  ${\bf P}_i^k$ at $(\theta_i^k, \phi_i^k)$ refers the surface point location on ellipsoid $\mathcal{E}_i$ ($i=1$ or 2) at iteration $k$,
and $k=0$ denotes the initial parameter values before the iteration starts.  
 Joining ${\bf P}_1^k$ and ${\bf P}_2^k$ by a line segment, we have the direction vector along the connecting line as
     \begin{equation}
     	{\bf d}_{12}=-{\bf d}_{21}={\bf P}_2^k - {\bf P}_1^k~~.
     	\label{eq:d12}
     \end{equation}
 One can visualize  the connecting line as an elastic string, and the tension force tends to pull the two end points together till the minimum-distance configuration is reached.
To mimic this effect, we project the direction vector ${\bf d}_{ij}$ on to the tangential directions
\begin{equation}
	\delta_{\theta,i}={\bf d}_{ij}\cdot{\bf e}_{\theta,i}~,~~~
	\delta_{\phi,i}={\bf d}_{ij}\cdot{\bf e}_{\phi,i}~~.
	\label{eq:delta_theta_phi}
\end{equation}  
Their relative magnitudes indicate in which direction we should move the surface point to reduce the distance. 
By introducing  an angle increment  step $\lambda_i$, we calculate the respective increment amounts $d\theta_i$ and $d\phi_i$ in the ${\bf e}_{\theta,i}$ and  ${\bf e}_{\phi,i}$ directions as
\begin{equation}
	d\theta_i=\frac{\delta_{\theta,i}}
	{(\delta_{\theta,i}^2+\delta_{\phi,i}^2)^{1/2}}\lambda_i~, ~~
	d\phi_i=\frac{\delta_{\phi,i}}
	{(\delta_{\theta,i}^2+\delta_{\phi,i}^2)^{1/2}}\lambda_i~~.
	\label{eq:lambda}
\end{equation}  
The new surface point location ${\bf P}_i^{k+1}$ is then given by
\begin{equation}
	\theta_i^{k+1}=\theta_i^k +	d\theta_i~, ~~
	\phi_i^{k+1}=\phi_i^k +	d\phi_i~~~.
	\label{eq:phi_new}
\end{equation}
The above described calculations should be performed twice, one to update ${\bf P}_1^k$ to ${\bf P}_1^{k+1}$ on ellipsoid $\mathcal{E}_1$ ($i=1$ and ${\bf d}_{ij}={\bf d}_{12}$),
and another
   one to update ${\bf P}_2^k$ to ${\bf P}_2^{k+1}$ on ellipsoid $\mathcal{E}_2$ ($i=2$ and ${\bf d}_{ij}={\bf d}_{21}$).
   We thus finish the iteration round, and this process can be repeated till the stopping criteria are satisfied, which will be discussed next.
   
\subsection{Implementation Details}
Other than the iteration calculations outlined above,
there are few implementation issues to be considered for adopting the surface-sliding method in practical computations:

~~~\\
{\bf Initial Locations of ${\bf P}_1^0$ and ${\bf P}_2^0$}:
The initial locations of the two surface points are necessary so we can start the first iteration. 
One option could be using the intersection points of the center-to-center connecting line to the two ellipsoidal surfaces. 
In particulate flow simulations and other similar applications,
the relative ellipsoid position and orientation are dynamically updated.
In such situation, the search  can start with the minimum-distance surface points at the previous simulation time step.
Since the change in particle configuration cannot be large over one time step, 
the search for the new minimum-distance locations should be fast.
In our numerical tests below, however, the initial positions of the surface points are assigned individually to challenge the algorithm robustness.

~~\\ 
{\bf Angle Increment Step $\lambda_i$}:
The angle increment step $\lambda_i$ in Eq. (\ref{eq:lambda}) plays a similar role as the search step in optimization \cite{numerical_book}.
In general, a large  $\lambda_i$ can speed up the search process; however, 
the calculation may become unstable and fail to find the minimum-distance points.
On the other hand, a small  $\lambda_i$ is helpful for numerical stability and robustness; but more iterations will be needed to reach the accuracy requirement.
In addition, 
strategies for adaptive step adjustment like in other numerical methods  \cite{numerical_book}
should be implemented to refine the step size in later stage of the search and thus a good accuracy can be achieved.
In our following demonstration examples, we set $\lambda_1^0=\lambda_2^0$ at the beginning and different values are tested.
As for the step refinement, we detect the overshoot situation if 
$|{\bf d}_{12}^k| > |{\bf d}_{12}^{k-1}|$, 
i.e., the distance increases in this iteration round $k$. 
When overshoot occurs, we halve $\lambda_1$ and $\lambda_2$ alternatively.
This simple strategy works well in our tests,
although more delicate methods can be considered \cite{opt_book}.

~~~\\
{\bf Stopping Criteria}:
Several quantities can be considered as candidates for the stopping criteria to terminate the iteration process.
The first one can be the relative error of the ${\bf P}_1 - {\bf P}_2$ distance:
\begin{equation}
	\epsilon_d=\left |
	\frac{|{\bf d}_{12}^k| - |{\bf d}_{12}^{k-1}|}{|{\bf d}_{12}^k|}
	\right |~~~.
	\label{eq:epsilon_d}
\end{equation}
For a converging search process, $\epsilon_d$ decreases and gradually approaches 0 at the minimum-distance locations. A threshold value can be set and the results at the end of the last iteration will be accepted as the solution of the problem.

Recall that the ${\bf P}_1$-${\bf P}_2$ connecting line should be aligned to the ellipsoid normal directions ${\bf n}_1$ and ${\bf n}_2$ when the minimum distance is found.
The angles between them can thus serve as an indicator for the closeness to the minimum-distance solution as well:
\begin{equation}
	\epsilon_n=\max\left(
	1-\hat{\bf d}_{12}\cdot {\bf n}_1~,~~~ 
	1+\hat{\bf d}_{12}\cdot {\bf n}_2 
	\right)~~, ~~~
	\hat{\bf d}_{12}=\frac{{\bf d}_{12}}{|{\bf d}_{12}|}~~~.
	\label{eq:epsilon_n}
\end{equation} 
This method has been used in the moving-ball method \cite{linhan2001}.

At last, the angle increment  step $\lambda_i$ can also be used to evaluate the converging process. 
As the search process approaches the minimum-distance solution, $\lambda_i$ is reduced gradually.
A small $\lambda_i$ value suggests that the current iteration has reached the close vicinity of the solution, and thus we can have 
\begin{equation}
	\epsilon_\lambda=\max \left( \lambda_1~,~~ \lambda_2\right)
\label{eq:epsilon_lambda}
\end{equation}    
as a criterion for terminating the calculation as well.
No specific stopping criterion is applied in our example calculations below;
however, the evolution of these converging indicators will be examined.

\begin{table}[]
	\caption{System parameters used for the demonstration examples in this research}
	\begin{tabular*}{\linewidth}{@{\extracolsep{\fill}}c c c c c c}
	\hline
		\multicolumn{2}{l}{Particle}    & 
		System I     & 
		System II - aligned   &
		System II - rotated   & 
		System III                     \\ 
		\multicolumn{2}{l}{Parameters}    & 
		(Fig. \ref{fig:converging_process}a)    & 
		(Fig. \ref{fig:rotation}a)  &
		(Fig. \ref{fig:rotation}b)  & 
		(Fig. \ref{fig:shape_size})                    \\ \hline
		\multirow{3}{*}{$\mathcal{E}_1$} & ($a_1, b_1, c_1$) & (1, 0.6, 0.4)                & (1, 0.6, 0.4)                & (1, 0.6, 0.4)                 & (0.2, 0.4, 0.6)               \\
		& ${\bf X}_{0,1}$    & (-1.5, 0, 0)                 & (-1.5, 0, 0)                 & (-1.0607, 0, -1.0607)         & (-1.0607, 0, -1.0607)         \\
		& ($\alpha_1, \beta_1, \gamma_1$)      & (0, $\pi/6$, 0) & (0, 0, 0)   
		             & (0, -$\pi/4$, 0) & (0, -$\pi/4$, 0) \\ \hline
		\multirow{3}{*}{$\mathcal{E}_2$} & ($a_2, b_2, c_2$)       & (0.6, 0.7, 0.5)    
		        & (1, 0.6, 0.4)                & (1, 0.6, 0.4)                 & see text \\
		& ${\bf X}_{0,2}$                 & (1, 0.5, 0.5)                & (1.5, 0, 0)                  & (1.0607, 0, 1.0607)           & (1.0607, 0, 1.0607)           \\
		& ($\alpha_2, \beta_2, \gamma_2$)        & (0, 0, $\pi/4$) & (0, $\pi/2$, 0) & 
		(0, $\pi/4$, 0)  & (0, $\pi/4$, 0)  \\ \hline
	\end{tabular*}
	\label{tbl:parameters}
\end{table}

\section{Demonstration Examples}
\label{sect:demos}
In this section, we present several demonstration examples with ellipsoids of various locations, orientations, sizes and aspect contrasts.
Different initial positions and search steps are also tested to investigate the algorithm robustness. 

\begin{figure}[thb]
	\centering
	\includegraphics[width=1\textwidth]{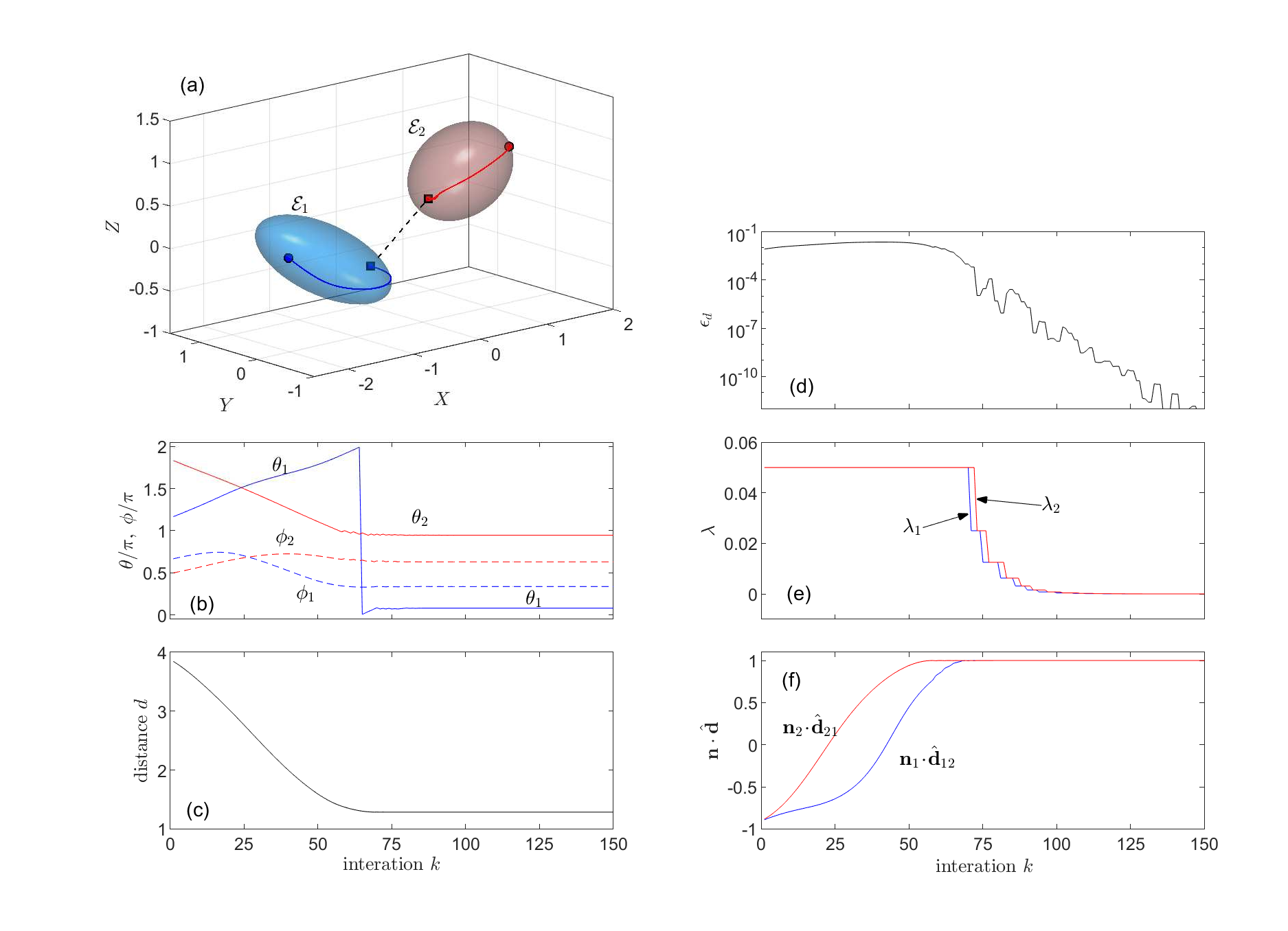}	
	\caption{The search process for System I in Table \ref{tbl:parameters}: 
	(a) a 3D view of the relative positions and orientations of the two ellipsoids,
	(b) the evolution of the surface point locations,
	(c) the converging process in distance between the surface points,
	(d) the relative error in separation distance,
	(e) the refining process of the angle increment steps, and
	(f) the improving alignment between the surface normal directions and the connecting segment.
	In (a), the circles represent the initial surface point positions and the squares for their final locations when the minimum distance is found, 
	and the curves over the ellipsoid surfaces are the surface point trajectories. 
	}
	\label{fig:converging_process}
\end{figure}

\subsection{Converging Process}
The first case  we test is the configuration shown in Fig. \ref{fig:converging_process}a.
The dimensions, positions and orientations of the two ellipsoids are listed in 
Table. \ref{tbl:parameters} (System I). 
The search starts with 
surface points at $(\theta_1^0, \phi_1^0)=( 7\pi/6,  2\pi/3)$ on $\mathcal{E}_1$ and
 $(\theta_2^0, \phi_2^0)=( 11\pi/6,  \pi/2)$ on $\mathcal{E}_2$.
These initial locations actually are on the opposite sides of 
those of the minimum-distance solution,
and this poses more challenge to the algorithm than using the center-to-center intersection points.
The angle increment steps are $\lambda_1^0=\lambda_2^0=0.05$.
The trajectories of the surface points during the search process can be visualized
 in Fig. \ref{fig:converging_process}a, where the circles represent the starting points and the squares 
 are the minimum-distance pair.
 More quantitatively, 
 Fig. \ref{fig:converging_process}b shows the evolution of surface point location parameters.
The sudden jump in $\theta_1$  occurs since, at iteration $k=65$, $\theta_1$ increases over  $2\pi$ and it is set back to the $0\sim 2\pi$ definition range of $\theta$.
It can be seen there that, once the search commences, all the position parameters start to change under the {\it tension} of the connecting line, and they reach their individual steady values quickly after $\sim75$ iterations.
In the meanwhile, the distance between the two surface points continuously decreases and a steady value is established at $d_{min}^*=1.2856$.
This value is considered as the minimum distance between the ellipsoids in Fig. \ref{fig:converging_process}a in our next analysis.

The converging process is further illustrated in Figs. \ref{fig:converging_process}d-f:
The relative distance error $\epsilon_d$ stays relatively constant in $0.01 \sim 0.02$ in the early stage 
$k=1\sim 60$;
however, it quickly drops later, to $<10^{-10}$ for $k>120$.  
Simultaneously, 
the refinement process for the angle increment steps $\lambda_1$ and $\lambda_2$ kicks off at $k=\sim 70$ 
as the surface points have moved to the vicinity of the minimum-distance solution.
The search steps reduce quickly to $\sim 10^{-6}$ at $k>120$, 
suggesting that our algorithm has successfully found the minimum-distance spots and it is just refining for a more accurate final result.  
Moreover, the dot products  
${\bf n}_i \cdot \hat{\bf d}_{ij}$
in Fig. \ref{fig:converging_process}f
quickly increase to 1, 
indicating that the connecting segment aligns nearly perfectly to the local normal directions 
and the correct minimum-distance situation is found.
Please note that this requirement is not directly involved in our search calculation, 
but automatically satisfied as the iteration converges.
Also their initial values are approximately -1, meaning the initial condition for our search algorithm is relatively harsh with the outward normals pointing in the opposite directions to 
the {\it tension} from the connecting segment.
This demonstrates the robustness of our algorithm in dealing with unfavorable search conditions.

\begin{figure}[thb]
	\centering
	\includegraphics[width=0.8\textwidth]{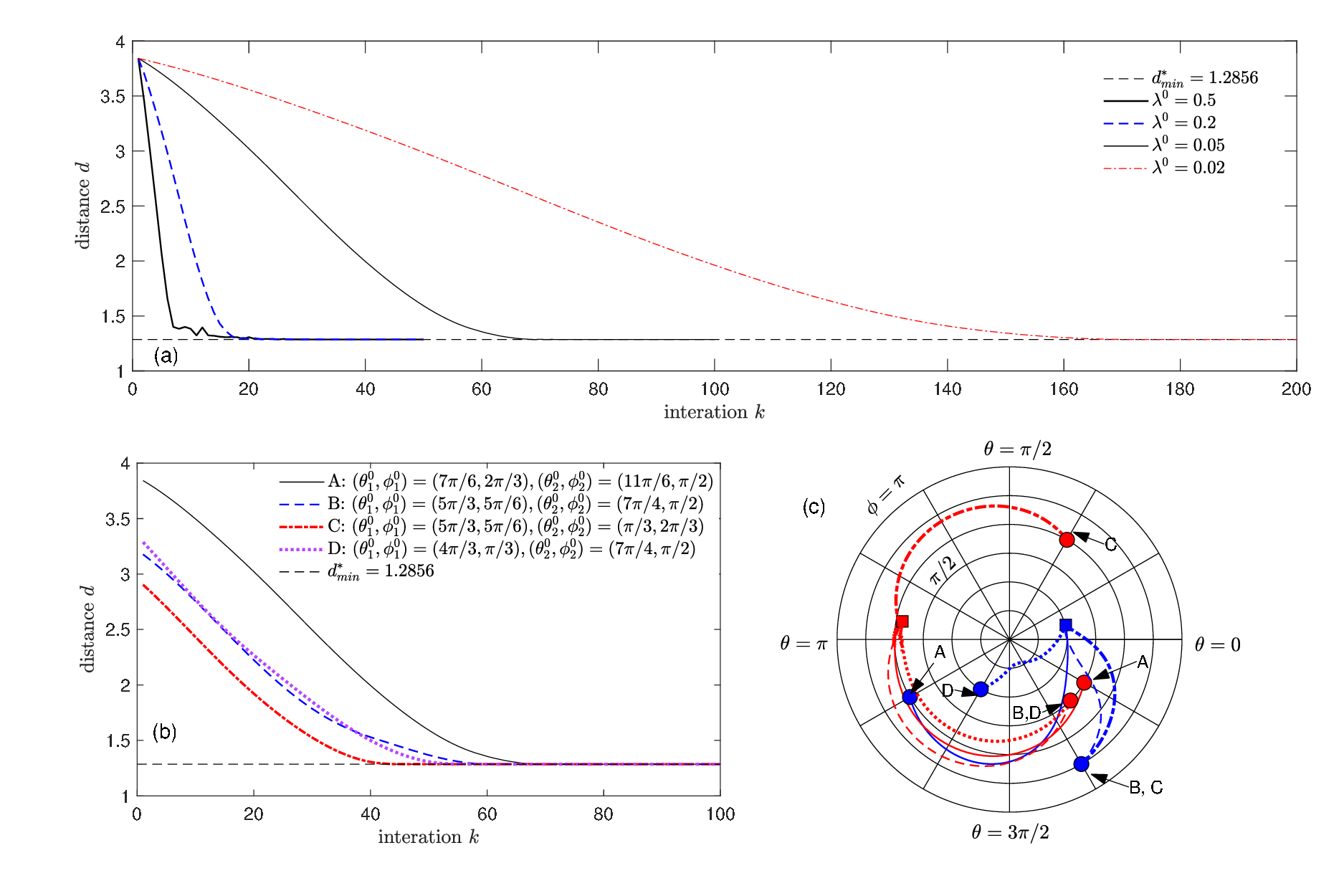}	
	\caption{
Influences of (a) different initial angle increment step $\lambda^0$
		and (b and c) different initial positions of the surface points. 
		In (c), the circles represent the initial surface point positions at the beginning of the search,
		 the squares represent their final positions when the solution is found,
		 and the curves between them indicate the search paths in the $\theta-\phi$ plane.		
		Different colors are used: blue for ellipsoid $\mathcal{E}_1$ and red for ellipsoid $\mathcal{E}_2$. 
		The same line styles are adopted in (b) and (c) for each case for clarity. 	
	}
	\label{fig:ini_conditions}
\end{figure}
   
\subsection{Initialization Insensitivity}
The same configuration in Fig. \ref{fig:converging_process}a is re-tested  in this subsection 
with different initial conditions.
Starting with the same initial positions  $(\theta_1^0, \phi_1^0)=( 7\pi/6,  2\pi/3)$ and
$(\theta_2^0, \phi_2^0)=( 11\pi/6,  \pi/2)$,
different values for the initial angle increment step $\lambda^0$ are utilized in Fig. \ref{fig:ini_conditions}a.
The observation is consistent with the general behaviors of search step influence in other numerical methods.
The large step value $\lambda^0=0.5$ causes oscillations at the beginning;
however, the search quickly converges.
All other $\lambda^0$ values yield smooth, monotonically decreasing curves for distance $d$.
It takes more iterations for small $\lambda^0$ values, which is reasonable since 
a smaller search step requires more iterations to approach the correct solution.
The minimum distance $d_{min}^*=1.2856$ is considered as the correct solution based on our analysis in the previous section, and it is plotted in Fig. \ref{fig:ini_conditions}a for comparison.
Clearly, all calculations converge to this value accurately, 
and this further confirms that  $d_{min}^*=1.2856$ is indeed the solution for the system considered here.
$\lambda^0=0.05$ will be used in all next calculations.

In addition, we test other three different sets of initial surface point locations,
denoted as Cases B-D in Fig. \ref{fig:ini_conditions}b with Case A for that used Figs. \ref{fig:converging_process} and \ref{fig:ini_conditions}a above.
The converging processes of separation distance $d$ is shown in Fig. \ref{fig:ini_conditions}b,
while their search paths are displayed in the $\theta-\phi$ plot in Fig. \ref{fig:ini_conditions}c.
The same line styles are adopted in Figs. \ref{fig:ini_conditions}b and c for clarity,
and the individual starting locations of these cases are labeled in Fig. \ref{fig:ini_conditions}c.
Although they start from different initial positions and follow different search routes, the same minimum-distance solution is found in $\sim 70$ iterations,
as indicated by the same converged distance $d_{min}^*=1.2856$ in Fig. \ref{fig:ini_conditions}b and
the same end points (squares) for all search paths in Fig. \ref{fig:ini_conditions}c.
It takes a little longer for Case A and a little shorter for Case C for converging to $d_{min}^*$,
since the separation distance $d$ is the largest at the beginning in Case A  (3.8427) and the least in Case C (2.9038).
With these results, we feel confident that the surface-sliding algorithm can handle different initial search conditions and find correct solutions.

 \begin{figure}[thb]
 	\centering
 	\includegraphics[width=0.8\textwidth]{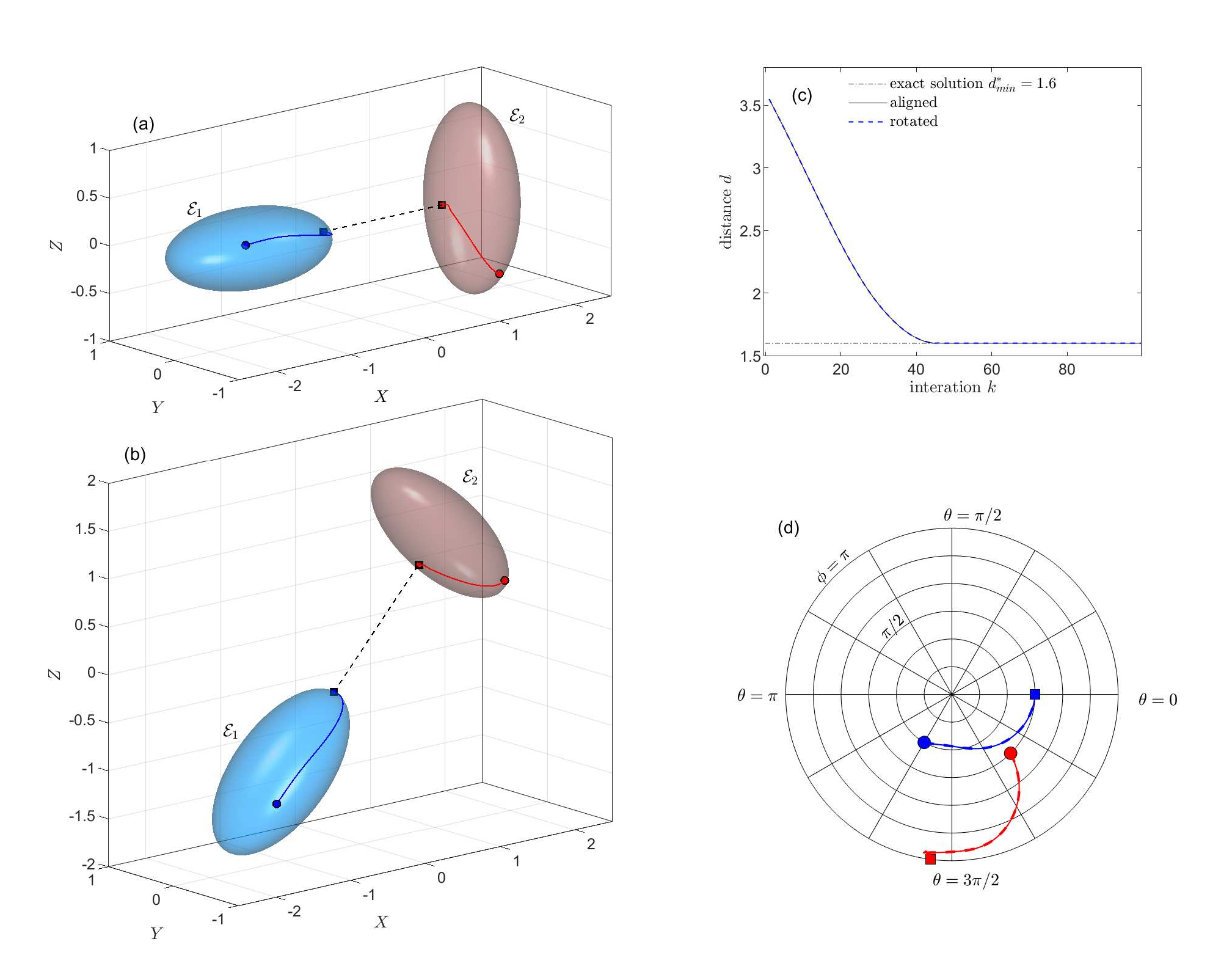}	
 	\caption{
 	3D views for System II in Table \ref{tbl:parameters} (a for the aligned configuration and b for the rotated configuration). 
 	The circles represent the initial surface point positions and the squares for their final locations when the minimum distance is found, 
 	and the curves over the ellipsoid surfaces are the surface point trajectories. 
 	 	The search processes with these two configurations are compared in (c) for the separation distance $d$ and (d) for the search paths in the $\theta-\phi$ plane. 
 	}
 	\label{fig:rotation}
 \end{figure}

\subsection{Spatial Isotropy}
The next comparison is carried out to test if the  surface-sliding method satisfies the spatial isotropy requirement, meaning that the minimum-distance solution should be  determined by the relative positions and orientations of the two ellipsoids, and independent from the artificial selection of the global coordinate system.
The two configurations considered are displayed in Figs. \ref{fig:rotation}a and b, and their respective parameters are given in Table \ref{tbl:parameters} (Systems II - aligned  and II - rotated).
It can be seen that the system  in Fig. \ref{fig:rotation}b is just a rotation of that in Fig. \ref{fig:rotation}a about the coordinate axis $Y$, and thus the same search process, including the converging path in the $\theta-\phi$ plot and the final minimum-distance solution, should be identical.
Also, for the particular setup in Fig. \ref{fig:rotation}a, 
the minimum distance occurs between the rightmost point on $\mathcal{E}_1$ at 
($\theta_1^*=0$, $\phi_1^*=\pi/2$) and
 the leftmost point on $\mathcal{E}_2$ at its south pole ($\phi_2^*=\pi$ and $\theta_2^*$ is not defined or arbitrary);
 and the minimum distance value 
is available analytically as 
\begin{equation}
	d_{min}^*=|{\bf X}_{0,2} -{\bf X}_{0,1}| -a_1 -c_2~~~.
	\label{eq:d_min}
\end{equation}      
With $|{\bf X}_{0,2} -{\bf X}_{0,1}| =3$, $a_1=1$ and $c_2=0.4$ for both system, 
we  thus have $d_{min}^*=1.4$.
The final search result, both the separation distance in Fig. \ref{fig:rotation}c 
and the surface point locations in Fig. \ref{fig:rotation}d, 
matches the above predictions perfectly.
Also, excellent agreement is noted in both the distance converging processes and the search paths in the $\theta-\phi$ plot for these two aligned and rotated setups, with no visual difference between the curves.
This comparison convincingly illustrates that our method can yield reliable results independent of the global coordinate frame selection.

 \begin{figure}[thb]
	\centering
	\includegraphics[width=1\textwidth]{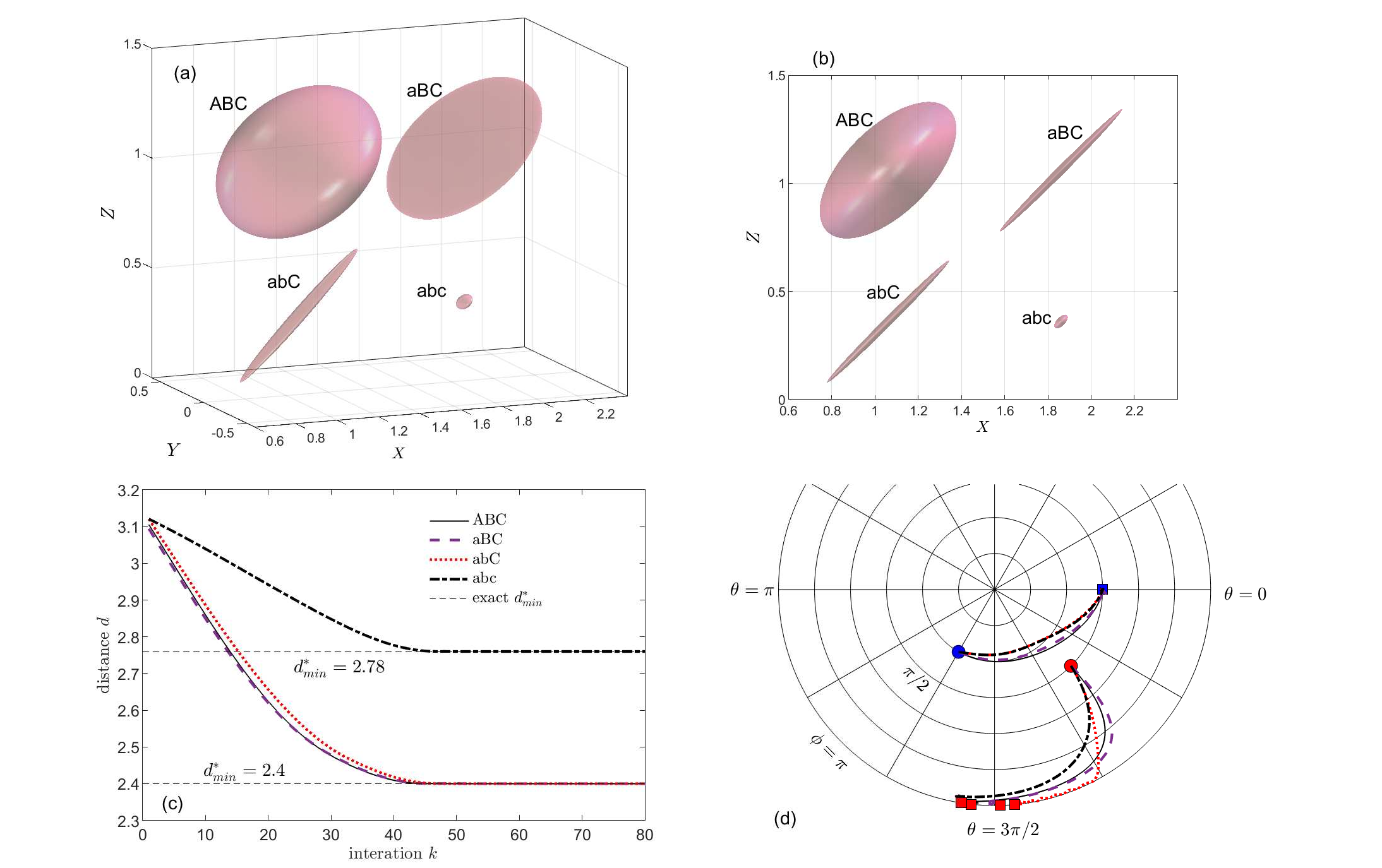}	
	\caption{
	Effects of ellipsoid size and geometry on the search process.
	The (a) 3D and (b) $X-Z$ views for the four different shapes of $\mathcal{E}_2$ are displayed together for comparison. 
	Detailed parameters are provided in Table \ref{tbl:parameters} (System III) and in the text.
	The converging processes in separation distance $d$ are given in (c) with the exact solutions  $d^*_{min}$ shown as dashed lines for comparison.
	The  search paths in the $\theta-\phi$ plane are also presented in (d), using the same line styles as in (c) for clarity. 
	}
	\label{fig:shape_size}
\end{figure}
   
\subsection{Robustness with Particle Size and Shape}
It has been reported that, for the moving-ball method, the convergence performance is less satisfactory for ellipsoids with large contrast in size and dimensions \cite{girault2022,linhan2001}.
To evaluate our algorithm on this aspect,
we use the rotated system in Fig. \ref{fig:rotation}b, however, 
with modifications on ellipsoid dimensions:
The $\mathcal{E}_1$ semi-axes are 
($a_1, b_1, c_1$)=(0.2, 0.4, 0.6), 
while four different  $\mathcal{E}_2$ dimensions are considered.
We denote the $\mathcal{E}_2$ shape identical to $\mathcal{E}_1$ with ($a_2, b_2, c_2$)=(0.2, 0.4, 0.6) as Shape ABC,
and label the other three shapes as:
Shape aBC with $(a_2, b_2, c_2)=({\bf 0.02}, 0.4, 0.6)$,
Shape abC with $(a_2, b_2, c_2)=({\bf 0.02}, {\bf 0.04}, 0.6)$,
and Shape abc with $(a_2, b_2, c_2)=({\bf 0.02}, {\bf 0.04}, {\bf 0.06})$.
From Shapes ABC to abc, we reduce one more semi-axis length tenfold every step. 
The reduced dimensions are highlighted in bold, and we use uppercase letters from the original length and the lowercase letters for the reduced dimension in the shape labels.
These shapes are compared in Figs. \ref{fig:shape_size}a and b, 
and the geometric contrast among them is distinctive.     
We can see there that Shape aBC represents a thin elliptical disk, 
Shape abC resembles a thin oblong blade,
and obviously Shape abc is just a much smaller version of Shape ABC.
The expression in Eq. (\ref{eq:d_min}) is still valid,
which yields $d_{min}^*=2.4$ for Shapes ABC, aBC and abC and 
$d_{min}^*=2.78$ for Shape abc.
The minimum-distance points are still 
    at  ($\theta_1^*=0$, $\phi_1^*=\pi/2$) on $\mathcal{E}_1$ and
    the south pole of $\mathcal{E}_2$ at $\phi_2^*=\pi$.
    
We start the search with the same initial positions 
($\theta_1^0=4\pi/3, \phi_1^0=\pi/3$) and 
($\theta_2^0=7\pi/4, \phi_2^0=\pi/2$), 
and use the same initial angle increment $\lambda^0=0.05$ for all the four $\mathcal{E}_2$ shapes.
The distance evolution and surface point trajectories during the search processes are displayed in
Figs. \ref{fig:shape_size}c and d, respectively.
For all cases, the correct minimum distance is quickly found in $\sim 50$ iterations, 
and the final surface point locations (squares in Fig. \ref{fig:shape_size}d) agree with our expectation excellently.
Please note that the four red squares in Fig. \ref{fig:shape_size}d actually represent the same search destination  on $\mathcal{E}_2$ - the south pole with $\phi_2^*=\pi$ and any arbitrary $\theta_2$ values.
Unlike in the moving-ball method \cite{girault2022,linhan2001}, the geometric aspect contrast has no apparent influence on the stability, accuracy and converging speed in our method, and there is no need to adjust the initial angle increment $\lambda^0$ for different ellipsoid dimensions.
This favorable feature is due to the fact that our search process actually proceeds in the $\theta-\phi$ space, which is not directly related to the ellipsoid size or shape.
On the contrary, the moving-ball method relies on the inner sphere radius to find the next surface location, and hence the ellipsoid size and local curvature matter for the search performance.

    \begin{figure}[thb]
    	\centering
    	\includegraphics[width=0.8\textwidth]{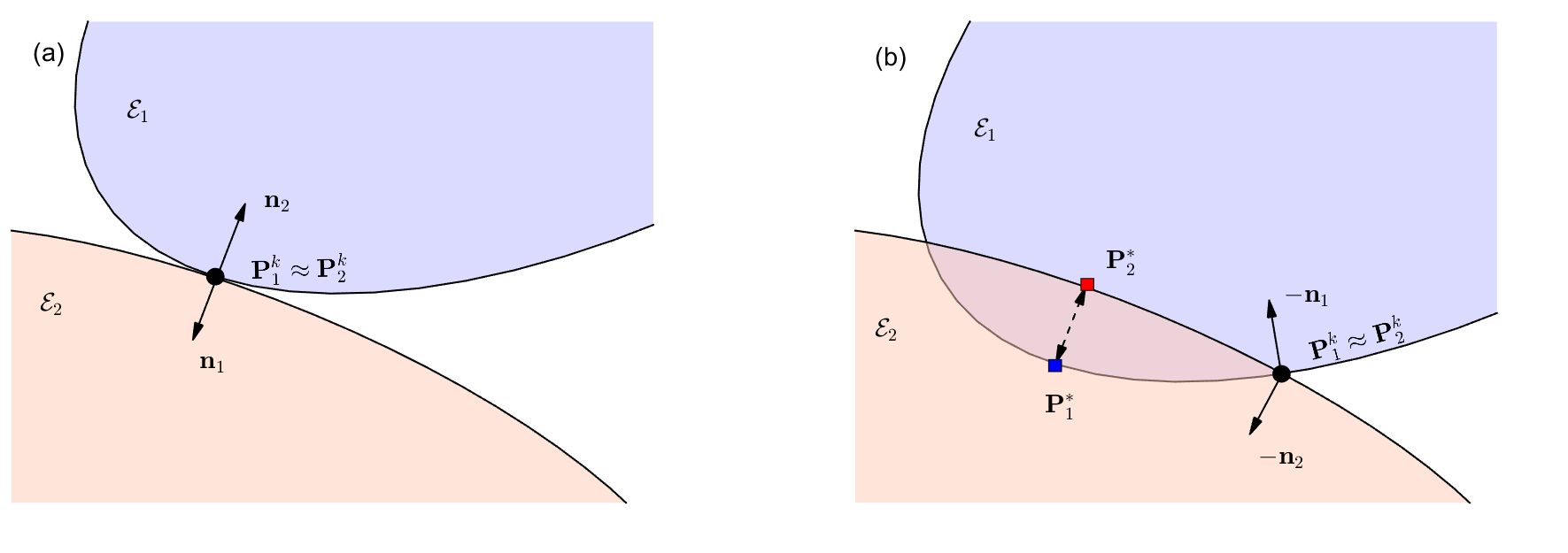}	
    	\caption{Two-dimensional illustrations for the (a) in-contact and (b) overlap situations of two ellipsoids.    	
    	}
    	\label{fig:contact}
    \end{figure} 

\section{Summary and Concluding Remarks}
\label{sect:summary}

We have proposed a novel surface-sliding method for finding the minimum distance and the corresponding surface point locations between two ellipsoids.
The iterative search process takes place in the individual $\theta-\phi$ space of each ellipsoid,
and the search direction is guided by the {\it tension} along the connecting segment.
Several demonstration examples have been designed to verify the numerical performance of our method, including the stability, accuracy and robustness.
Compared to other existing techniques,
our method  possesses unique features such as the clear geometric concept and 
simple mathematical formulation.
This surface-sliding method is expected to be an attractive choice for computer graphics, engineering design and computational sciences.

In the method description and numerical examples, we have not considered the situations when the two ellipsoids are in direct contact (Fig. \ref{fig:contact}a) or they may even overlap each other (Fig. \ref{fig:contact}b). 
These situations need to be addressed in some applications such as the discrete element model and material modeling \cite{Banna_2025,You_2018_DEM,YZS_2025}.
This concern can be conveniently incorporated in our method as follows.
At the end of iteration $k$, if the distance $|{\bf P}_2^k-{\bf P}_1^k|$ is smaller than a threshold value  
$\sigma$, one can consider that the two ellipsoids are either in contact or intersecting each other.
By comparing ${\bf n}_1 \cdot {\bf n}_2$ versus -1, 
we can separate the in-contact (if ${\bf n}_1 \cdot {\bf n}_2 \approx -1$) condition from the overlap situation.
No further treatment is required for the in-contact case.
If  the overlap  (penetration) magnitude (dashed double arrow in Fig. \ref{fig:contact}b) is  needed, we can continue the search 
from ${\bf P}_1^k$ under the direction of $-{\bf n}_2$ on $\mathcal{E}_1$, 
and similarly from ${\bf P}_2^k$ under the direction of $-{\bf n}_1$ on $\mathcal{E}_2$.
Once the distance increases back beyond the threshold $\sigma$, we can resume the regular update 
calculation, and gradually approach the maximum overlap locations ${\bf P}_1^*$ and ${\bf P}_2^*$ (Fig. \ref{fig:contact}b). 
At last, we  would also like to mention that this surface-sliding concept can be readily applied 
to other scenarios, 
such as the minimum distance problems between an ellipsoid and a fixed location or a flat wall \cite{You_2018_DEM, computer_graphics},
between other parametric surfaces \cite{Sohn_2002},
and between a convex particle and the concave wall in a container.
These extensions are straightforward, and researchers with such needs can quickly modify the corresponding equations we present above to accomplish their individual tasks.  
 
\section*{Acknowledgments}
This work was supported by the Natural Science and Engineering Research Council of Canada (NSERC). 

 \bibliographystyle{ieeetr}
 
\bibliography{draft}

\end{document}